\documentclass[aip,pop,article-title,reprint]{revtex4-1}
\usepackage{hyperref}
\usepackage{graphicx}
\usepackage{amsmath}
\usepackage{amssymb}
\usepackage{bm}

\def\Wp{W_\parallel}
\begin{document}
\title{Synthetic Multidimensional Plasma Electron Hole Equilibria}
\author{I H Hutchinson}
\email{ihutch@mit.edu}
\affiliation{Plasma Science and Fusion Center\\
Massachusetts Institute of Technology\\
Cambridge, MA 02139, USA}
\begin{abstract}
  Methods for constructing synthetic multidimensional electron hole
  equilibria without using particle simulation are
  investigated. Previous approaches have various limitations and
  approximations that make them unsuitable within the context of
  expected velocity diffusion near the trapped-passing boundary.  An
  adjustable model of the distribution function is introduced that
  avoids unphysical singularities there, and yet is sufficiently
  tractable analytically to enable prescription of the potential
  spatial profiles. It is shown why simple models of the charge
  density as being a function only of potential cannot give solitary
  multidimensional electron holes, in contradiction of prior
  suppositions. Fully self-consistent axisymmetric electron holes in
  the drift-kinetic limit of electron motion (negligible gyro-radius) are
  constructed and their properties relevant to observational
  interpretation and finite-gyro-radius theory are discussed.
\end{abstract}
\maketitle

\section{Introduction}
Active space plasma regions are observed to contain long-lived
solitary positive potential peaks whose spatial extent is a few Debye
lengths; they are mostly identified as electron holes, having an
electron charge deficit on trapped
orbits\cite{Matsumoto1994,Ergun1998,Bale1998,Mangeney1999,Pickett2008,Andersson2009,Wilson2010,Malaspina2013,Malaspina2014,Vasko2015,Mozer2016,Hutchinson2018b,Mozer2018}.
It has been known for a long time that non-zero magnetic field is
necessary for the sustainment of electron holes; if it is strong
enough, the electron motion and trapping becomes
one-dimensional. One-dimensional dynamic treatments have predominated
past hole theory, where the gyro-radius $r_L$ is neglected relative to
the transverse scale length $L_\perp$. Although the present theory
continues to calculate using only parallel particle dynamics, it takes
account qualitatively of one recently discovered important effect of
the transverse electric field in a multidimensional electron hole,
namely the resonant interaction of the trapped particle bounce motion
with the gyro-motion. This interaction essentially always induces a
region of stochastic orbits near zero parallel energy: the
trapped-passing boundary of phase space. And the energy-depth of this
stochastic layer increases with
$r_L/L_\perp$\cite{Hutchinson2020}. The anticipated result in the
stochastic layer is a large effective phase-space diffusion rate which
forces the distribution function (phase-space density) to be
approximately independent of energy in that region. It has also been shown
recently\cite{Hutchinson2021}, in contradiction of a longstanding
suggestion, that regardless of the relative strength of the magnetic
field, the screening of the trapped electron deficit charge is
isotropic: having approximately Boltzmann dependence. This discredits
one speculation concerning how the transverse scale of electron holes
relates to magnetic field strength, and reemphasizes the need to
understand better multidimensional electron holes, especially since
recent multi-satellite measurements are giving unprecedented
information about multidimensional holes in space plasmas.

\begin{figure}
  \centering
  \includegraphics[width=0.9\hsize]{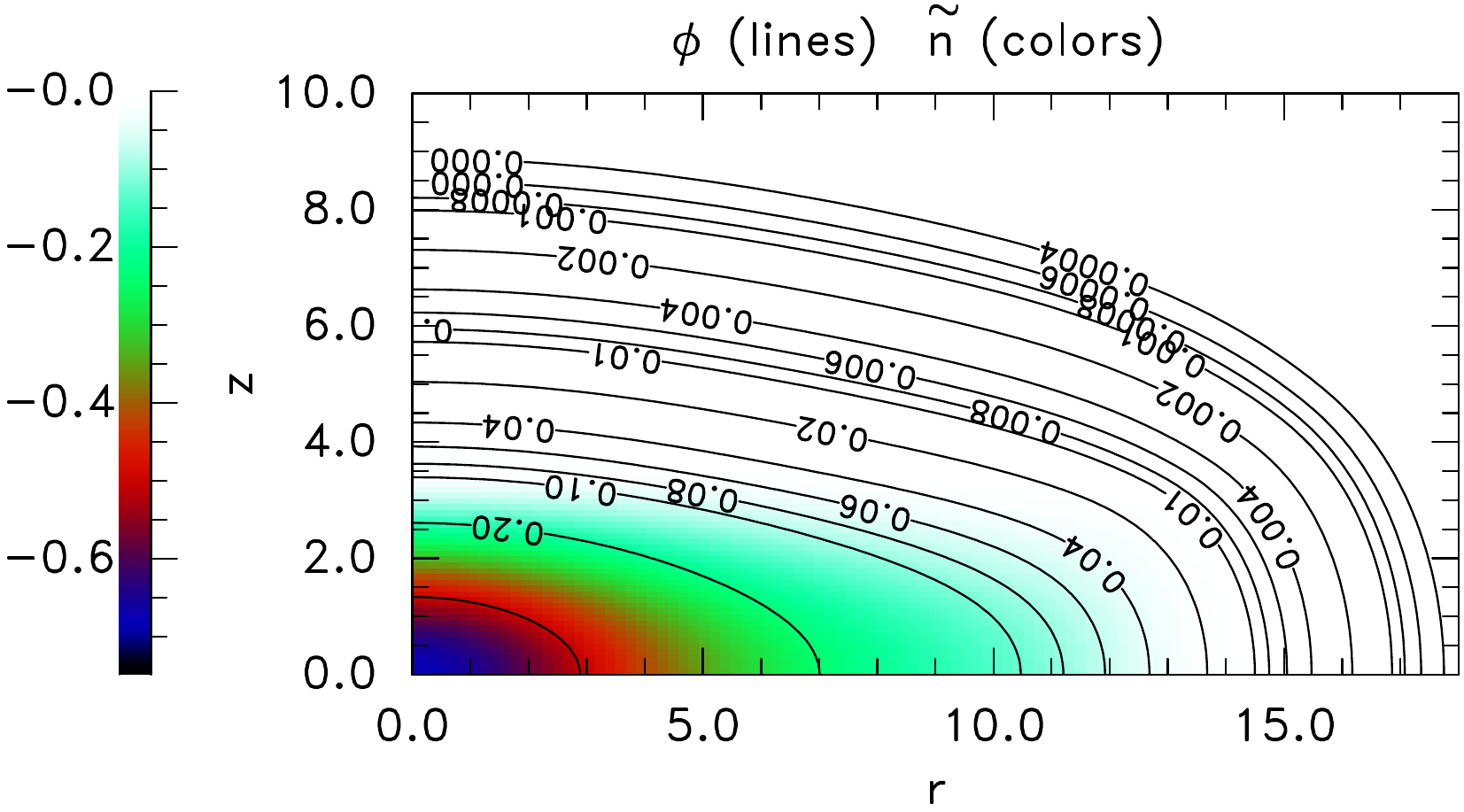}
  \caption{An oblate electron hole equilibrium, axisymmetric about the
    magnetic field direction $z$, showing color contours of the
    trapped electron density deficit $\tilde n$, and line contours of
    the self-consistent potential $\phi$.}
  \label{fig:onex}
\end{figure}

The purpose of the present work is to develop a versatile model of
self-consistent multidimensional electron hole equilibria that conform
to the physical constraint of having zero distribution slope at and
for a controllable depth inside the trapped-passing boundary.  Because
of the difficulties of taking fully into account finite gyro-radius
dynamic effects, the model simply prescribes the parallel velocity
distribution as a function of transverse position $r$, and potential
$\phi$. This exactly represents the electron dynamics only for the
limit of high magnetic field (one-dimensional motion) where the
gyro-radius is small. However, even this first step has not previously
been achieved taking properly into account the full Poisson equation
and resulting non-separable form of its solution for the potential.
The present results provide physically
self-consistent potential profiles in which the consequences of
transverse dynamics, and especially gyro-averaging, will be explored in
a separate publication.  An illustrative equilibrium found by the
present approach (explained fully later) is shown in Fig.\
\ref{fig:onex}.

After a review of prior analysis of multidimensional electron hole
equilibria (section 2), a model distribution function with
shape controlled by adjustable parameters is introduced in section
3. It avoids the unphysical properties of most previous choices, and
from it explicit analytic expressions for the charge deficit and (for some
parameters) the one-dimensional potential profile can be derived. It
is generalized to separable multidimensional potential forms. In
section 4 is explained the subtlety of how to relax the potential
iteratively to the non-separable form it always must take. It becomes
clear that no solitary multidimensional equilibrium can in fact exist
when the charge density is a function only of potential. It must be
also an explicit function of transverse position. Iterative
relaxation with effectively constant trapped charge deficit embodies
sufficient of the physical stabilizing dynamics to produce a
convergent scheme. And examples of hole shapes are examined
with particular emphasis on the consequences for satellite
observations.

For convenience throughout this paper normalized units are used:
Debye-length $\lambda_{D}$ for length, inverse plasma frequency
$1/\omega_{pe}$ for time, and background temperature $T_{e\infty}$ for
energy; with the result that velocity units are
$\sqrt{T_{e\infty}/m_e}$. Densities are normalized to the distant
background electron density $n_{e\infty}$. In steady state, the total
energy $W=v^2/2-\phi$ (normalized units) of an electron is conserved,
and neglecting collisions, the total derivative of the distribution
function $df(v)/dt$ along particle orbits is zero. Ions are taken to be a
uniform immobile neutralizing background.

\section{Two dimensional hole equilibria}

Electron holes are a type of BGK equilibrium\cite{Bernstein1957} in
which a deficit of the electron velocity distribution on trapped
orbits sustains the positive potential that traps them. For decades
electron hole analysis was almost all one-dimensional, along the
magnetic field and ignored or highly simplified any transverse spatial
dependence. This one-dimensional prior theory has been reviewed
extensively
elsewhere\cite{Turikov1984,Schamel1986a,Eliasson2006,Hutchinson2017}
and will not be discussed in detail here, but a brief review of the
relatively few \emph{multidimensional} electron hole analytic
equilibrium studies is appropriate.

Multidimensional electrostatic hole model studies were published for
several years following 2000, motivated by new observational
data documenting finite transverse hole extent in space plasmas. The
model equilibria based their particle dynamics on the one-dimensional
Vlasov equation plus drift in a uniform magnetic field, but began to
account for transverse potential variation, generally simplifying
the problem to axisymmetric cases (independent of $\theta$ in
cylindrical coordinates, which is our focus here too). 

\citet{Chen2002} specified the potential to have a separable
axisymmetric form and a constant value of $\nabla_\perp^2\phi/\phi$ ,
giving $\phi=\phi_\parallel(z) J_0(l_0r/r_s)$ ( the Bessel function
$J_0$ has its first zero at argument $l_0$). The potential is thus
zero, but has finite $\nabla_\perp$, at $r=r_s$. This represents a
``waveguide'' configuration, effectively equivalent to the (1979) analysis of
Schamel\cite{Schamel1979}, rather than an isolated hole far from boundaries. The
convenience of the ansatz is that the Laplacian in Poisson's equation
simply acquires from transverse divergence an extra term
$k_\perp^2\phi$ where $k_\perp^2=(l_0/r_s)^2$ is independent of
$r$. This term is readily included into an expression for charge
density in Poisson's equation. With potential specified in this way,
and an assumed Maxwellian untrapped distribution, the authors solve
the integral equation for the trapped parallel distribution function
$f_\parallel(W_\parallel)$ (for parallel energy including potential
$W_\parallel<0$). One commonly discussed criterion is that
$f_\parallel$ must be non-negative, which constrains the hole parallel
length in a way that depends on the perpendicular scale $r_s$.

Later, \citet{Chen2004} analyzed instead Gaussian potential shapes
$\phi\propto\exp(-z^2/\delta_z^2-r^2/\delta_\perp^2)$ (still separable
but far from boundaries), showing that the required trapped density as
a function of potential, and trapped distribution as a function of
energy, can be derived in closed form for Maxwellian background
distribution. They also included the static response of a repelled
(ion) Maxwellian species of different temperature.

\citet{Muschietti2002} prescribed a separable potential
$\phi \propto [1+\eta\cosh(\beta z)]^{-1}\exp[-(r/\delta_\perp)^2]$,
which allows parallel elongation of the hole by flattening its top via
the parameter $\eta$, and prescribing its asymptotic scale-length
$1/\beta$ at large $z$. Its perpendicular variation is Gaussian, which
means that writing $\nabla_\perp^2\phi(z)=k_\perp^2\phi$, implies
$k_\perp^2$ varies with $r$.  By making a mathematically convenient
(non-Maxwellian) choice of the external parallel distribution, they
found a closed analytic form for the trapped distribution, satisfying
the parallel integral equation self-consistently as a function of
energy and radius $r$. This calculation (like those of Chen et al) is
carried out in the drift kinetic approximation, assuming the ordering
$\omega_\theta\ll \omega_b\ll \Omega_e$ where $\omega_\theta$ is the
azimuthal drift frequency of the gyrocenter about the axis, $\omega_b$
is the parallel bounce frequency, and $\Omega_e$ the cyclotron
frequency. It also ignores the distinction between guiding-center density
and particle density, effectively neglecting the gyro radius relative
to the perpendicular scale length. The work explores only limited
transverse extent $r/\delta_\perp\le 1.3$.

These three early papers using the ``integral equation'' solution
approach (prescribing the parallel potential shape) all have a slope
singularity in the parallel distribution function at the
separatrix where the parallel energy
$W_\parallel$ is zero,
$f_\parallel(W_\parallel)-f_\parallel(0)\propto
(-W_\parallel/\psi)^{1/2}$, which Muschietti et al illustrate. And
they all use separable potential form.

\citet{Jovanovic2002a} approached the problem
instead by the ``differential equation'' route, specifying the trapped
parallel distribution function to be of the Schamel form
($\propto\exp(\beta W_\parallel/T)$), which for small potential gives
a density difference from pure Debye shielding (i.e.\
$n_e/n_0=1+\phi$) proportional to $-\phi^{3/2}$, and no
$f_\parallel$-slope singularity. They supposed that the Poisson
equation was modified by anisotropic dielectric shielding, becoming
effectively
$\{\nabla_\parallel^2+[1+(\omega_p/\Omega_e)^2\nabla_\perp^2]\}\phi=a\phi-(4b/3)\phi^{3/2}$.
This supposed modification of Poisson's equation is erroneous when
applied to electron holes, as has been shown elsewhere\cite{Hutchinson2021}.

\citet{Krasovsky2004a} show that in a plasma with (Debye) shielding
length $\lambda_D$, a spherically symmetric hole of Gaussian form
$\phi=\psi\exp(-R^2/\lambda^2)$ is self-consistent with a trapped
parallel distribution function deficit
$\tilde f_\parallel=f_{\parallel
  trapped}(W_\parallel)-f_\parallel(0)=-(2|W_\parallel|/\pi^2)^{1/2}[1/\lambda_D^2
+(2/\lambda^2)(1+2\ln(4|W_\parallel|/\psi) ]$, governed by the Vlasov
equation. For mathematical convenience, they invoke instead the
presumption that the trapped particle density (deficit), like the
shielding density, depends linearly on the potential, which is
satisfied by having
$\tilde f_\parallel \propto \sqrt{W_j-W_\parallel}$ for
$W_\parallel/W_j>1$ and zero for $W_\parallel/W_j<1$, where
$W_j=-\phi_j$ is some maximum trapped energy ($W_j<0$). The advantage
is that the linear dependence of density gives rise to Helmholtz's
equation $(\nabla^2+k^2)(\phi-const.)=0$ except with the constant
$k^2=-dn/d\phi$ different (in magnitude and sign) in the two regions
$W_\parallel/W_j\lessgtr 1$. The potential in the inner region can be
expressed in terms of a sum of known harmonic solutions, by specifying
the position in $r$ and $z$ of the potential contour at the join
$\phi=\phi_j$.

In \citet{Krasovsky2004} these authors venture beyond the pure
drift approximation, noting that a potential of the \emph{additive}
form $\phi=\psi\times[1-(z/L)^2-(r/R)^2]$ gives rise to integrable
equations of motion. Although they do not find an equilibrium, they
note that the resulting (2-D) Vlasov equation is satisfied by $f(v)$
being an arbitrary function of the energy-like constants of the
motion, $w_\parallel = v_z^2/2 +(z/L)^2\psi$, and
$w_\perp = v_r^2/2+(r/R)^2\psi+(p_\theta/r+Br/2)^2/2$, where
$p_\theta$ is the conserved canonical angular momentum about the axis of
symmetry in the magnetic field $B$, and the total energy is actually
$W=w_\perp+w_\parallel-\psi$. They concentrate on the density at
$r=0$, $z=0$, noting that orbits that pass through the origin have
$p_\theta=0$, and are trapped only if they never reach $\phi=0$, which
is true if $W_\parallel + 1/[1+(BR)^2/8\psi]W_\perp<\psi$. This
inequality defines the interior of a trapped ellipse in velocity space
at the origin, and thereby limits the maximum possible charge deficit
contributable by trapped particles. To make that exceed the passing
particle density perturbation (which at a minimum it must), when the
field is weak $B\lesssim 1$ and $\psi\ll B$, they show requires
$(r_L/R)^2\sqrt{\psi}\lesssim 1$. Thus, a maximum thermal gyro
radius $r_{Le}$ of order the hole's radial scale length ($R$) times
$\psi^{-1/4}$ is permitted. A similar criterion is derived by
Krasovsky et al (2006).

After 2006 I am aware of no published analytic assaults on the
multi-dimensional hole equilibrium problem until
\citet{Hutchinson2020} showed that when transverse potential gradients
exist there is always a region of stochastic orbits caused by
bounce-cyclotron resonance near $W_\parallel=0$. The energy depth of
the stochastic region is found as a function of the peak potential
$\psi$, $E_\perp$, and the magnetic field strength $B$. It deepens
rapidly when $B/\sqrt{\psi}\lesssim 2$ and
$E_\perp$ is significant, eventually extending to the full hole
depth. The presence of this stochastic region will prevent the
formation of any strong energy gradients of $f_\parallel$ near
$W_\parallel=0$, ruling out any distributions that do not have an
approxmately flat region of $f_\parallel$ there. As has been noted
before\cite{Hutchinson2017}, this forces a requirement that the hole
potential fall $\propto \exp(-z/\lambda_D)$ at large z, but it also
constrains $\phi \propto \exp(-r/\lambda_D)$ at large $r$. Thus, exact
Gaussian potential forms, parallel or perpendicular, cannot be
physical in the hole wings.

These summaries motivate the key emphases of the present work:
insisting upon physically plausible distribution dependence on
energy, and abandoning the convenient but unphysical assumption
that the potential form is separable. Observing these principles we
construct truly physical multidimensional electron hole equilibria,
based for the first time on synthetic analysis rather than
particle-in-cell simulation.

\section{Power trapped distribution deficit model}

Consider a solitary potential peak which in its own frame of
reference is time-independent. The distribution function satisfying
the (parallel) Vlasov equation is constant on particle orbits. Since
the potential is steady, energy is also constant on orbits, and in the
drift approximation in a uniform magnetic field the distribution is a
function of parallel ($W_\parallel=v_\parallel^2/2-\phi$) and total
($W$) energy, with those energies conserved.

\begin{figure}
  \centering
  \includegraphics[width=0.9\hsize]{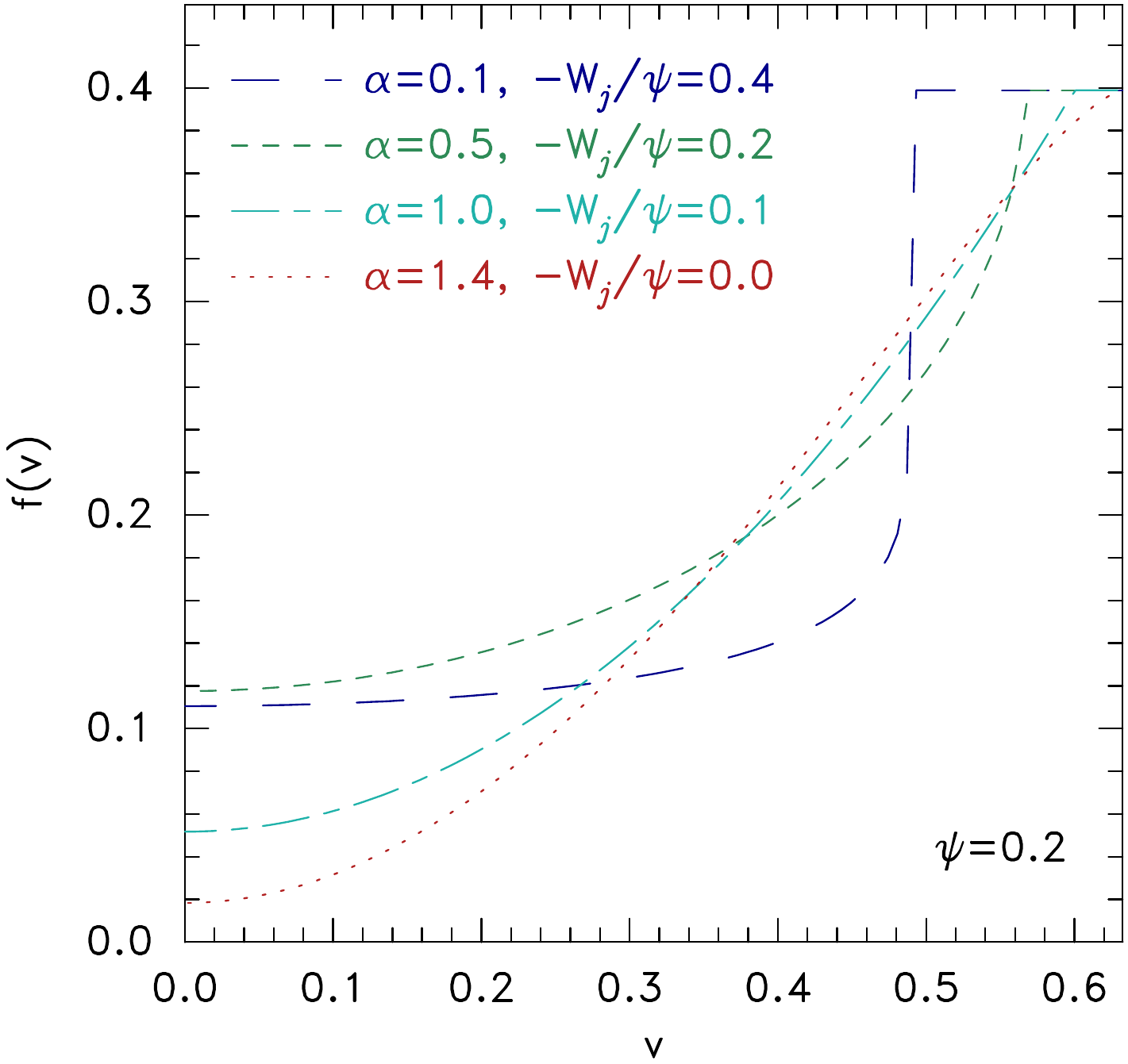}
  \caption{Illustration of shapes of the distribution function in the
    trapped region of velocity space at potential $\psi$, for the
    power deficit model.}
  \label{fig:powerex}
\end{figure}

However, for orbits that are weakly trapped, conservation of magnetic
moment (and hence of $W_\parallel$ separately from
$W=W_\parallel+W_\perp$) begins to break down because of bounce-gyro
resonance\cite{Hutchinson2020}, and resulting velocity space diffusion suppresses the
difference $\tilde f =f_t-f_\infty(0)$ between the trapped
distribution and a flat distribution having the separatrix value
$f_\infty(0)$. We therefore adopt initially the ``differential equation''
($\tilde f$ specified) approach to constructing a self consistent
equilibrium solution. We suppose, further, that there is a bounding
trapped energy $W_j$, lying between $-\psi$ and $0$, such that for
parallel energy $\Wp >W_j$, $\tilde f$ is zero. For the more deeply
trapped region where $\Wp <W_j$, $\tilde f$ is taken proportional to
the energy difference raised to a chosen power $\alpha$:
$\tilde f \propto(W_j-\Wp )^\alpha$. This form is convenient because
it is continuous and analytically tractable, yet enables monotonic
shapes of $\tilde f(\Wp )$ to be represented from uniform waterbag
($\alpha=0$), through rounded $\alpha \sim 1/2$ to triangular
$\alpha=1$, and beyond, peaked at the maximum trapping depth. It is here called
the power deficit model. We take no account of the perpendicular
velocity distribution and so $f$ means $f_\parallel$ and we drop the
parallel suffix for brevity henceforth.
Fig.\ \ref{fig:powerex} illustrates some of the distribution function shapes that can be
prescribed by the power deficit $\tilde f$ model. 

\subsection{One-dimensional}
If the overall magnitude of $\tilde f$ is represented by its value
$\tilde f_\phi$ at parallel energy $-\phi$ for some specific $\phi>-W_j$ (i.e.\
in the non-zero region), then its value at all other energies
($\Wp <W_j$) is
\begin{equation}
  \label{eq:tildef}
  \tilde f(\Wp ) = \left(W_j-\Wp \over \phi +W_j\right)^\alpha\tilde f_\phi
  = C (W_j-\Wp )^\alpha,
\end{equation}
where $C\equiv \tilde f_\phi/(\phi+W_j)^\alpha$ is a constant
(negative because $\tilde f$ is negative).
We adopt for brevity hereafter the convention that energy combinations
taken to some real power give zero if they are negative. 
The density perturbation produced by $\tilde f$ at potential $\phi$ is
given by the integral
\begin{equation}
  \label{eq:tilden}
  \begin{split}
  \tilde n(\phi) &= \int_0^{\sqrt{\phi}} \tilde f\, 2dv
  = \int_{-\phi}^{W_j} \tilde f_\phi\left(W_j-\Wp \over \phi
    +W_j\right)^\alpha { \sqrt{2}\, d\Wp \over\sqrt{\phi+\Wp }}\\
  &= 2\tilde f_\phi(\phi +W_j)^{1/2}
  \sqrt{2}\int_0^1(1-\zeta^2)^\alpha d\zeta\\
  &=2CG (\phi+W_j)^{\alpha+1/2}
  \end{split}
\end{equation}
[using the substitution $\zeta\equiv\sqrt{(\phi+\Wp )/(\phi+W_j)}$, and
writing $G$ for the $\zeta$-integral (which is a known function of
$\alpha$)
$ G \equiv \sqrt{2}\int_0^1(1-\zeta^2)^\alpha d\zeta =
\sqrt{\pi/2}\,\Gamma(\alpha+1)/\Gamma(\alpha+3/2)$].

Denote the densities due to trapped electrons $n_t$, passing electrons
$n_p$, and passing plus a flat distribution of trapped electrons
$n_f$. The charge-density $\rho=1-n_p-n_t=1-n_f-\tilde n$ is a
function of $\phi$, so Poisson's equation in one-dimension
$d^2\phi/dz^2=-\rho$ can be integrated once to give
\begin{equation}
  \label{eq:Poisson1}
  {1\over 2}\left(d\phi\over dz\right)^2=\int(n_f-1)+\tilde n d\phi\equiv-V_f-\tilde V=-V_{total}.
\end{equation}
For our power deficit model $\tilde f$ we have
$-\tilde V = 2CG(\phi+W_j)^{\alpha+3/2}/(\alpha+3/2)$.  The flat-$f$
density for a Maxwellian at small $\phi$ is $n_f=1+\phi$, giving (the
``classical potential'') $-V_f=\phi^2/2$, and $\tilde n$ can be
evaluated at larger $\phi$ or for other distributions, e.g.\ shifted
Maxwellians, numerically. In any case, the implicit solution for the
potential form is then $z(\phi)=\int d\phi/\sqrt{-2V_{total}}$. The
crucial condition at the hole peak, $z=0$, where $d\phi/dz=0$, is then
$V_{total}=0$. The peak potential $\psi=\phi(0)$ is thus the solution
of
\begin{equation}
  \label{eq:Vpeak}
\begin{split}
  -\tilde f_\psi {2G(\psi+W_j)^{3/2}\over \alpha+3/2}&=
  -2CG{(\psi+W_j)^{\alpha+3/2}\over \alpha+3/2}\\ &=\tilde V(\psi)=
  -V_f(\psi).
  \end{split}
\end{equation}
For the \emph{linear} $n_f$ approximation $V_f=-\phi^2/2$, this equation gives
\begin{equation}
  \label{eq:Ceq}
  \tilde f_\psi/\psi^{1/2}=-{(\alpha+3/2)\over 4G} {1\over(1+W_j/\psi)^{3/2}},
\end{equation}
which shows that $\tilde f_\psi \propto \sqrt{\psi}$ with a constant of
proportionality that depends only on the $\tilde f$ shape parameters
$\alpha$ and $W_j/\psi$.
Also for the linear $n_f$ approximation
\begin{equation}
  \label{eq:CG}
  CG=-{(\alpha+3/2)\over 4} {\psi^{1/2-\alpha}\over(1+W_j/\psi)^{3/2+\alpha}};
\end{equation}
so
\begin{equation}
  \label{eq:ntilde}
  \tilde n(\phi)= -{(\alpha+3/2)\over 2} {\psi^{2}(\phi+W_j)^{1/2+\alpha}
    \over(\psi+W_j)^{3/2+\alpha}}
  .
\end{equation}

There does not appear to be a closed form solution for the $z(\phi)$
integral for general $\alpha$. However, as was by Krasovsky et
al\cite{Krasovsky2004a}, there is when one adopts the linear $n_f$
approximation and $\alpha=1/2$, because the resulting linear
dependence of the density on $\phi$ turns Poisson's equation into the
Helmholtz equation for $\phi>-W_j$ and the Modified Helmholtz equation
for $\phi <-W_j$, which match at the join.
\begin{equation}
  \label{eq:helmholtz}
  \begin{array}{lll}\displaystyle
    {d^2\phi\over dz^2}&= \phi+2 C G(\phi+W_j)\\
                       &=-k^2(\phi-2CGW_j/k^2)\qquad&(-\phi<W_j)\\
    &= \phi &(-\phi>W_j).
  \end{array}
\end{equation}
The inner region solution is an offset cosine:
\begin{equation}
  \label{eq:cosform}
\phi-2CGW_j/k^2=(\psi-2CGW_j/k^2)\cos(kz)  
\end{equation}
where $k^2=-(1+2CG)$, and the
amplitude comes from requiring $\phi(0)=\psi$. The outer
region is an exponential
\begin{equation}
  \label{eq:expform}
\phi=-W_j\exp(z_j-z).  
\end{equation}
The join position is
where
$\cos(kz_j)=[k^2+2GC]/[k^2\psi/W_j+2CG]=W_j/[k^2\psi+2CGW_j]$. For
$\alpha=1/2$ the value of $G$ is $\pi/(2\sqrt{2})$, and 
\begin{equation}
  \label{eq:Clin}
  C={\tilde f_\psi\over(\psi+W_j)^{1/2}}= {\sqrt{2}\over \pi} {1\over(1+W_j/\psi)^{2}}.
\end{equation}

We note that there are certain constraints on allowable values of the
parameters of the power deficit model. One is that the distribution
function cannot be negative, so $|\tilde f|\le
f_\infty(0)=1/\sqrt{2\pi}=0.399$
(unshifted Maxwellian); this limits
the allowable maximum $\psi$ to a value of order unity. Another
less obvious is that for $\alpha<1/2$ there is a minimum allowable
value of $-W_j/\psi$ for there to exist a solution to $V(\psi)=0$, and
when $V_f=-\phi^2/2$ it is $-W_j/\psi > 1/4-\alpha/2$. The minimum
value $1/4$ for a waterbag $\tilde f$ (i.e.\ $\alpha\to 0$) was noted long
ago by \citet{Dupree1982}.
\begin{figure}
  \centering
  \includegraphics[width=0.8\hsize]{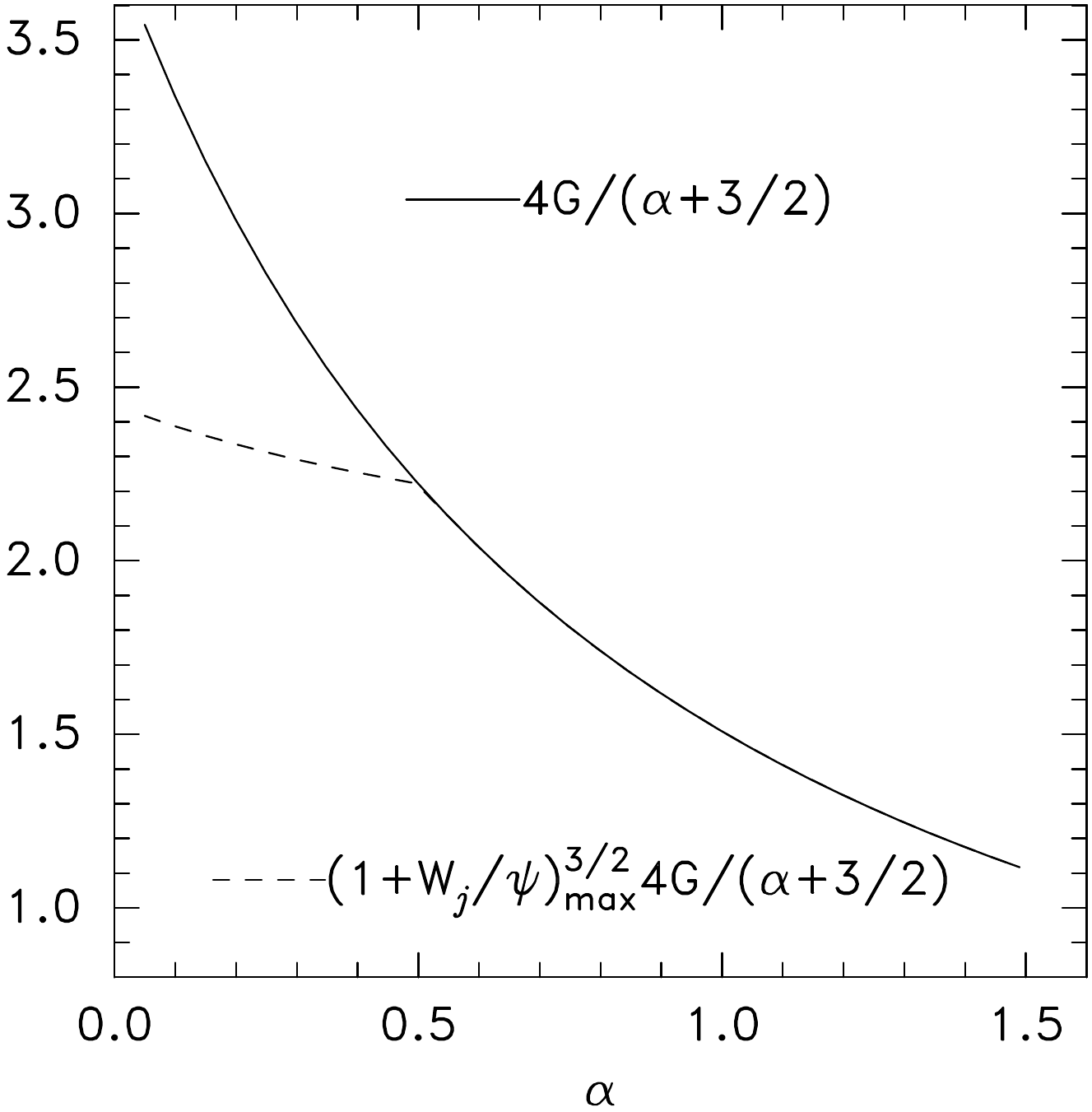}
  \caption{Dependence of the key coefficient on power $\alpha$.\label{fig:fmaxcoef}}
\end{figure}
Fig.\ \ref{fig:fmaxcoef} shows how the numerical coefficient $4G/(\alpha+3/2)$
varies with $\alpha$. Taking into account the allowable minimum of
$-W_j/\psi$ the maximum value of  $(1+W_j/\psi)4G/(\alpha+3/2)$ does not
rise as steeply for $\alpha<0.5$. This combination is what determines
the maximum value of $\psi^{1/2}$ for a given $\tilde f_\psi$ through
eq.\ \ref{eq:Ceq}, and hence how the non-negativity constraint varies
with $\alpha$.

So far this is all one-dimensional analysis.

\subsection{Multidimensional Generalization: Separable Potential-form}
When there is transverse potential gradient, Poisson's equation
contains an additional transverse divergence term $\nabla_\perp^2\phi$
that couples together Vlasov solutions on adjacent field-lines. If its
value at a certain transverse position $r$ is considered a known
function of parallel position $z$, then the integration along $z$ can
still be carried out and the transverse divergence adds an extra term
$V_\perp = \int \nabla_\perp^2\phi\, d\phi$, giving classical
potential $V_{total}=\tilde V + V_f +V_\perp$. In the separable case
where $\nabla_\perp^2\phi=\phi/L^2$, with $L(r)$ the transverse
scale-length independent of $z$, the effective density contribution is
again proportional to $\phi$, and $V_\perp= \phi^2/2L^2$.  Writing
$-2[V_f(\psi)+V_\perp(\psi)]/\psi^2=F_\perp$, equations
(\ref{eq:Ceq}), (\ref{eq:CG}), and (\ref{eq:ntilde}), coming from
$V_{total}=0$, can be generalized by multiplying the right hand sides
by $F_\perp$ as
\begin{equation}
  \label{eq:Ceqm}
  \tilde f_\psi/\psi^{1/2}=-{(\alpha+3/2)\over 4G}
  {1\over(1+W_j/\psi)^{3/2}}F_\perp;
\end{equation}

\begin{equation}
  \label{eq:CGm}
  CG=-{(\alpha+3/2)\over 4}
  {\psi^{1/2-\alpha}\over(1+W_j/\psi)^{3/2+\alpha}}F_\perp;
\end{equation}
and
\begin{equation}
  \label{eq:ntildem}
  \tilde n(\phi)= -{(\alpha+3/2)\over 2}
  {\psi^{2}(\phi+W_j)^{1/2+\alpha}  \over(\psi+W_j)^{3/2+\alpha}}F_\perp.
\end{equation}
These expressions provide the trapped deficit distribution
$\tilde f(W_\parallel,r)$ required for a specified transverse
potential variation $\psi(r)$ and join energy $W_j(r)$, \emph{if the
  potential were separable}. However, adopting that distribution
\emph{does not make the potential separable}. In fact $\phi$ is
\emph{never} exactly separable, because (at least) far from the hole
(where $\tilde n$ is negligible) it becomes a Yukawa potential, which
is not separable. Therefore it must be emphasized that synthetic
expressions (\ref{eq:cosform}) and (\ref{eq:expform}) for potential
proportional to $\cos(kz)$ ($\phi>-W_j$) and $\exp(-z)$ ($\phi<-W_j$),
or other separable power deficit potential model approximations, are
not fully self-consistent equilibria. Fortunately, the approximation
can be quite good, but demonstrating so requires us to find the exact
self consistent potential $\phi(r,z)$ corresponding to this
distribution using a numerical solution of the Poisson system with the
corresponding $\tilde n(\phi,r)$.

\begin{figure}[htp]
  \centering
  \includegraphics[width=0.9\hsize]{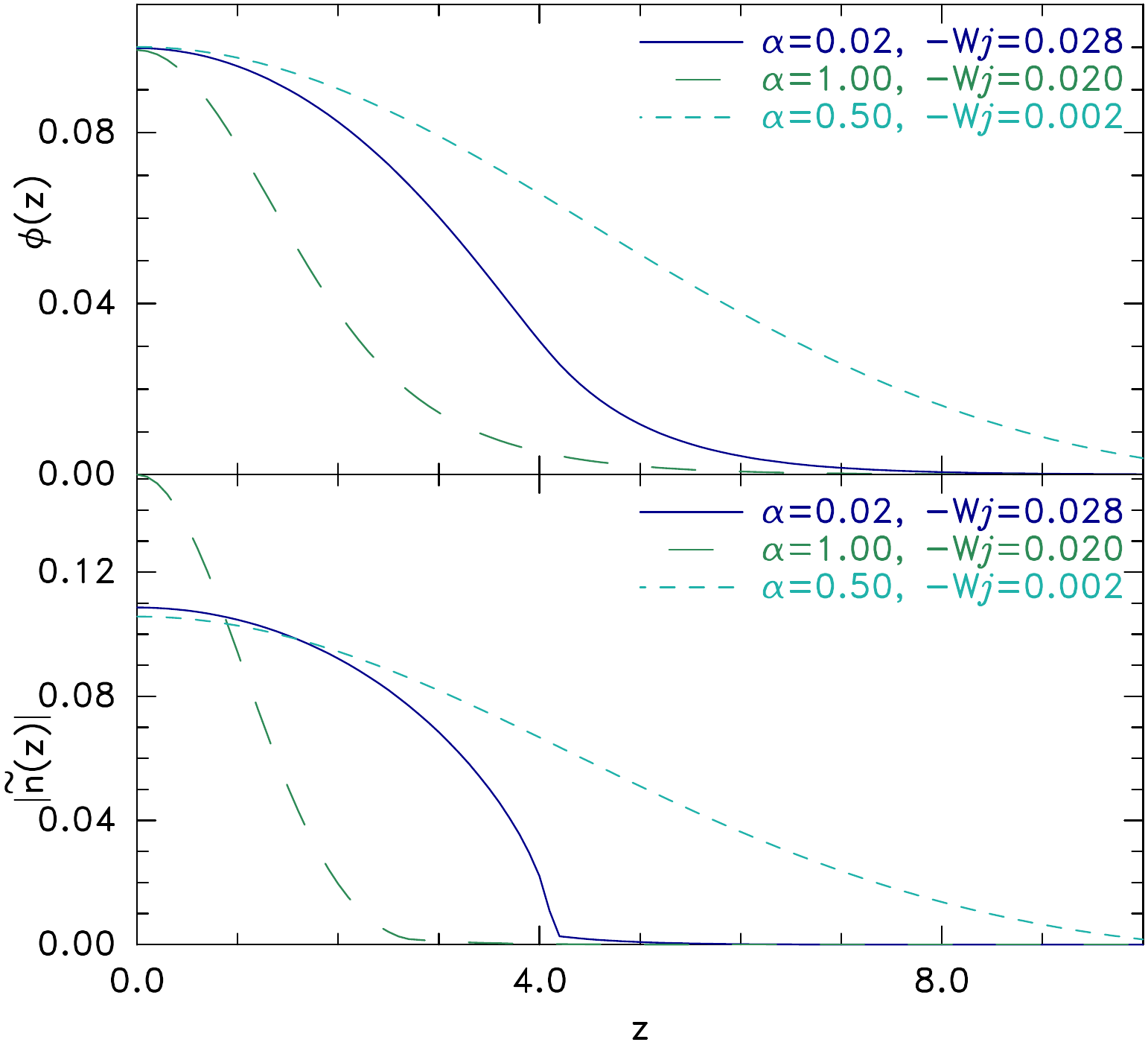}
  \caption{Potential and density deficit $z$-profiles (at $r=0$),
    illustrating different model parameter settings.}
  \label{fig:zprofiles}
\end{figure}
Different types of parallel ($z$) profiles are illustrated in Fig.\
\ref{fig:zprofiles}. They are actually profiles at $r=0$ of fully
self-consistent multidimensional solutions, whose attainment is the
subject of the next section, but they are little different from the
separable treatment of this section.  When the power is very small,
$\alpha=0.02$, the $\tilde f$ shape is essentially a waterbag (flat
for $W<-W_j$, and zero for $W>W_j$) giving a density deficit
$|\tilde n(z)|$ that is nearly elliptical. This case gives the
steepest onset of deficit, and has a $\tilde f$-slope singularity, but
at $W=W_j$, not the separatrix.\footnote{Its barely-observable
  positive value of $|\tilde n|$ beyond $z=4$ $\phi=-E_j$ corresponds
  to the small difference between $n_f$ and $1+\phi$, which is added
  for the purpose of solving the Modified Helmholtz equation (in other
  words the quantity plotted is, strictly speaking,
  $|\tilde n + n_f -(1+\phi)|$). This correction is thus relatively
  unimportant for these cases where $\psi=0.1$ and the hole speed is
  zero.} The potential has only moderate curvature at $z=0$.

By contrast, when $\alpha=1$ the $\tilde f(W)$ is linear, without a
slope singularity anywhere, but naturally tends to peak more near
$z=0$, giving charge and potential profiles that are considerably
narrower. The case $\alpha=0.5$ makes $|\tilde n|$ a linear function
of $\phi$; and choosing a small $-W_j=0.002$ ($-W_j/\psi=0.02$) causes the
profiles to extend to considerably larger $z$ than the other
examples, even though the $\phi$-curvature at $z=0$ is almost the same as
the waterbag case. A limitation of this power deficit model is that it
cannot represent the non-monotonic $\tilde f(W)$ profiles required to
give a flat region of $\phi(z)$ near $z=0$.

\section{Relaxation to true non-separable equilibrium}

In addition to providing analytic self-consistent solutions when
potential gradients are ignorable or separable, the model potential is
also suitable for calculating numerically the full non-separable
potential form for two and three dimensional potential
variation. Obviously, to produce a hole of finite transverse dimension
one must have transverse (e.g.\ radial) variation of the peak
potential $\psi$, which will imply for the present model radial
variation of $\tilde f_\psi$ and $W_j$ (and possibly $\alpha$ but here
we shall suppose $\alpha$ is uniform).

In principle it ought to be possible to construct self-consistent
profiles for \emph{any} \emph{specified distribution function} of the
form $\tilde f =\tilde f(\Wp,r)$ which gives rise to charge-density
$\rho(\phi,r)=1-n_p-n_t=1-n_f-\tilde n= 1-n(\phi,r)$.  When
substituted into Poisson's equation $ -\nabla^2\phi=\rho(\phi,r)$, a
well-posed problem arises with some boundary conditions. It is
non-linear in general, of course, so typically will require numerical
solution, although in regions where $\tilde n$ is zero, the Modified
Helmholtz equation (Debye shielding) in multiple dimensions will be
obtained when $n_f$ is linear in $\phi$.

Particle-in-cell (PIC) simulations have frequently observed
multi-dimensional holes arising from initially uniform unstable plasma
distributions\cite{Mottez1997,Miyake1998a,Goldman1999,Oppenheim1999,Muschietti2000,Oppenheim2001b,Singh2001,Lu2008}. And isolated
equilibria depending on non-uniform trapped distributions have been
constructed and demonstrated to persist for extended time durations
using a PIC code by the present author. But a PIC code uses more
physics than just a knowledge of $n(\phi,r)$.  It turns out that to
find a multidimensional potential structure that is \emph{solitary}
(having boundary conditions only at infinity), based purely on
$\rho(\phi)$, is in fact \emph{not} a well posed problem.  The work
of \citet{Krasovsky2004a} using the two-region piecewise
linear-density variation $\rho(\phi)$ (as discussed above, with
$\alpha=1/2$ but without any explicit dependence on $r$) purports to
find a non-spherical equilibrium with a universal $\rho(\phi)$
dependence by prescribing a non-spherical contour shape of the join
potential $\phi_j=-W_j$, solving the resulting Helmholtz equation for
$\phi$ inside the contour as a sum of harmonics, and thereby
prescribing the fixed $\tilde n(r,z)$. However, their presumption that
this solution can be extended without discontinuity to the external
region appears to be incorrect. In any case --- and this is the key
problem of any multidimensional solitary equilibrium --- a
$\rho(\phi)$ non-spherical equilibrium prescribed by a $\phi_j$
contour shape constraint is unstable to potential changes if the
constraint is removed.

A numerical iterative relaxation on a suitable multidimensional mesh
of a potential profile in accordance with $-\nabla^2\phi=\rho(\phi,r)$
demonstrates this instability.  Explorations of such relaxation
approaches have shown that schemes based upon iterations in which
Poisson's equation is solved for given $\rho$ followed by updating $\rho$ in
accordance with prescribed $\rho(\phi,r)$ do not generally converge
to electron holes. The potential either grows without
limit in the large-$\phi$ region, or collapses to zero giving a null
($\phi=0$ everywhere) converged solution. And this behavior is
independent of the degree of partial relaxation, because it is not
merely numerical. The instability is physically intuitive, since
$\rho$ is an increasing function of $\phi$ in the positive charge
region, and positive $\rho$ generally gives rise to positive $\phi$ in
Poisson's equation, which amounts to positive feedback of a
perturbation near the potential peak. 

In one dimension, it is sometimes possible to stabilize relaxation
iterations so as to converge to the desired equilibrium by using an
implicit scheme, but this requires the iteration to start close enough
to the final potential profile, otherwise either collapse or
stochastic fluctuations result. Normally of course one-dimensional
equilibria are obtained for specified $\rho(\phi)$ not by relaxation
but by direct integration using the ``classical'' potential
$V(\phi)=\int \rho d\phi$. Such an approach does not readily carry
over to multi-dimensional problems, because the transverse field
divergence contribution is known only when the potential profile is
known.  In multiple dimensions, even implicit relaxation schemes
experience instability.

Physically steady, stable, solitary electron hole equilibria
nevertheless exist. How? The answer is that their actual
time-dependent electron dynamics is \emph{not represented by
  prescribing $\rho(\phi,r)$}. The most important dynamic effect is probably
that a time rate of change of the potential violates the conservation
of total energy along particle orbits, thereby dynamically changing
the form of $\rho(\phi,r)$ in response to a non-steady perturbation. A
successful numerical relaxation scheme to find the equilibrium must
represent some of that stabilizing physical dynamics. One way to
represent the dynamics approximately is to regard the charge density
as being specified as a function of the potential difference between
the peak potential $\psi$ (at $z=0$) and the potential elsewhere
($z\not=0$) so that $\rho=\rho(\psi-\phi,r)$. This mocks up the idea that
$\rho$ does not change in response to time-dependent evolution of
$\phi$ at only a fixed point, but rather has non-local dynamic response
contributions. Raising or lowering uniformly the entire potential
profile does not then change $\rho(z)$.

I have implemented schemes that take just the trapped particle deficit
$\tilde f$ to be a fixed function of $\psi-\phi$ (and $r$). The
screening response, consisting of the passing particles at positive
energy and flat trapped distribution $f_f$ at negative energy (which
in total I call the flat-trap or reference distribution) remains, as
before, a fixed function of $\phi$ (approximately $\propto\phi$). This
amounts to supposing that the passing particle density plus the flat
contribution in the trapped region responds quickly to changes in
potential, while the trapped \emph{deficit} does not. Heuristic
justification is that passing particles are rapidly exchanged out of
the hole and the flat trap level $f_\infty(0)$ does not depend on
$\phi$. Screening is a stabilizing contribution,
approximately the Boltzmann response. The power model of density
deficit is used; so mathematically the density derived from eq.\
(\ref{eq:ntildem}) is
\begin{equation}\label{eq:ntdiff}
  \tilde n(\phi)= -{(\alpha+3/2)\over 2}
  {\psi_d^{2}(\phi-\psi+\psi_d+W_{jd})^{1/2+\alpha}  \over(\psi_d+W_{jd})^{3/2+\alpha}}F_\perp,
\end{equation}
where $\psi_d$, $W_{jd}$, and $F_\perp$ (and $\alpha$) are fixed
quantities expressing the desired (subscript $d$) $\psi$, $W_j$,
initialized before the relaxation, and the $F_\perp$ value calculated
from the $\psi_d$ profile $V_{\perp d}$ plus $V_f$.  The $\phi$ and
$\psi$ change with iteration, but by prescription $\tilde n$ depends only on
their difference.

Two different
algorithms for solving for the potential update have been
investigated, one using an ad hoc implicit advance along alternating
directions. The other (much faster) simply solves Poisson's
equation as a Modified Helmholtz equation in an $rz$-domain, given the
difference from pure linear screening (the non-Helmholtz part of the
charge density, $\Delta\rho$ using the prior $\phi$) as a source:
\begin{equation}
  \label{eq:helmpois}
  \nabla^2\phi-\phi/\lambda_s^2 = -\Delta\rho
  = \tilde n +n_f-1-\phi/\lambda_s^2,
\end{equation}
where the screening length is taken as $\lambda_s=(dn_f/d\phi)^{-1/2}$
at $\phi\to 0$. The Helmholtz equation (\ref{eq:helmpois}) is solved at
each iteration using the cyclic reduction routine
\verb!sepeli! from the ``Fishpack'' library. Boundary conditions are
$d\phi/dr=0$ at $r=0$, $d\phi/dz=0$ at $z=0$,
$d\ln\phi/dr=-1/\lambda_s-1/r_{max}$ at $r=r_{max}$, and
$d\ln\phi/dz=-1/\lambda_s$ at $z=z_{max}$.
The distant boundaries do not matter much as long as they are far enough
away. These two codes use in addition a simple Pad\'e approximation
for the fully nonlinear screening (flat-trap) response $n_f$ of an
external Maxwellian of arbitrary drift velocity (not just the
$n_f-1= \phi$ approximation). They thus accommodate
any speed or depth of hole. The agreement between the two codes helps
verify their coding.

\subsection{Global density functional yields only 1-D holes}

One version of these relaxation calculations supposes that the
potential difference $\psi-\phi$ on which the electron density deficit
$\tilde n$ depends, is the quantity
$\phi(0,0)-\phi(r,z)$. That is, the difference between the local
potential and the potential at the origin, which is where the global
peak in potential lies. This may be called the ``global functional''. The
other, called the ``parallel functional'', is instead
$\phi(r,0)-\phi(r,z)$; that is, the difference at fixed radial
position between the local potential and the ridge at $z=0$ along the
parallel coordinate $z$.

A summary of the results of the global functional model relaxation is
simple. Electron hole equilibria are found, but they are spherically
symmetric, no matter what the initial starting state is. Even highly
elongated initial potential shapes rapidly relax to {spherically
  symmetric} equilibria that depend solely on the spherical radial
distance from the origin $R=\sqrt{r^2+z^2}$. These are therefore
one-dimensional, not multidimensional, though the one dimension is
spherical radius not a cartesian coordinate.

\subsection{Radially varying density parallel functional gives 2-D holes}

In this section we examine the two-dimensional (axisymmetric) holes
obtained when the density deficit takes the form of the power model,
eq.\ (\ref{eq:ntdiff}), in which $\psi=\phi(r,0)$, that is, the parallel
functional model $\tilde n$. The profile $\psi_d(r)$ is effectively a free
choice but the results shown will use the following parametrization
whose monotonic shape is sufficiently versatile for present purposes:
\begin{equation}
  \label{eq:psid}
  \begin{split}
  &\psi_d(r)=\psi_{d0}{1+\exp(r_t/\lambda_s)\over
    \cosh(r/\lambda_s)+ (1+Dr^2)\exp(r_t/\lambda_s)},\\
  &W_{jd}(r)=W_{j0}\psi_d(r)/\psi_{d0}.
  \end{split}
\end{equation}
Here the adjustable (uniform) parameter $r_t$, when positive, is
approximately the radius of a flattened potential region for
$r\lesssim r_t$; large negative $r_t$ removes all flattening, giving a
${\rm sech}(r/\lambda_s)$ shape. Parameter $D$ is approximately half an
extra negative curvature $-(1/\phi)d^2\phi/dr^2$ of the potential,
modifying the top's flatness if desired. The uniform screening length
$\lambda_s$ is not considered to be adjustable, but depends on the
passing particle distribution, notably its mean speed relative to the
hole, or equivalently the hole speed. The form chosen for eq.\
(\ref{eq:psid}) recognizes that at large $r$, where $\tilde n$ is
negligible, the transverse potential variation (at $z=0$) is dominated
by the screening length and asymptotes to
$1/\cosh(r/\lambda_s)\to\exp(-r/\lambda_s)$. The join energy is taken
to have a desired profile the same shape as $\psi_d$, giving a
constant desired ratio $W_{jd}(r)/\psi_d(r)=W_{j0}/\psi_{d0}$, with
values in the approximate range $-0.05$ to $-0.5$ being physically
plausible for most $\alpha$ values. Recall that, for
$W_\parallel>W_j$, the trapped distribution function is independent of
energy, equal to $f_0(0)$, plausibly accommodating rapid trapping and
detrapping of particles on stochastic orbits with energies above
$W_j$.
\begin{figure}
  \centering
  \includegraphics[width=0.8\hsize]{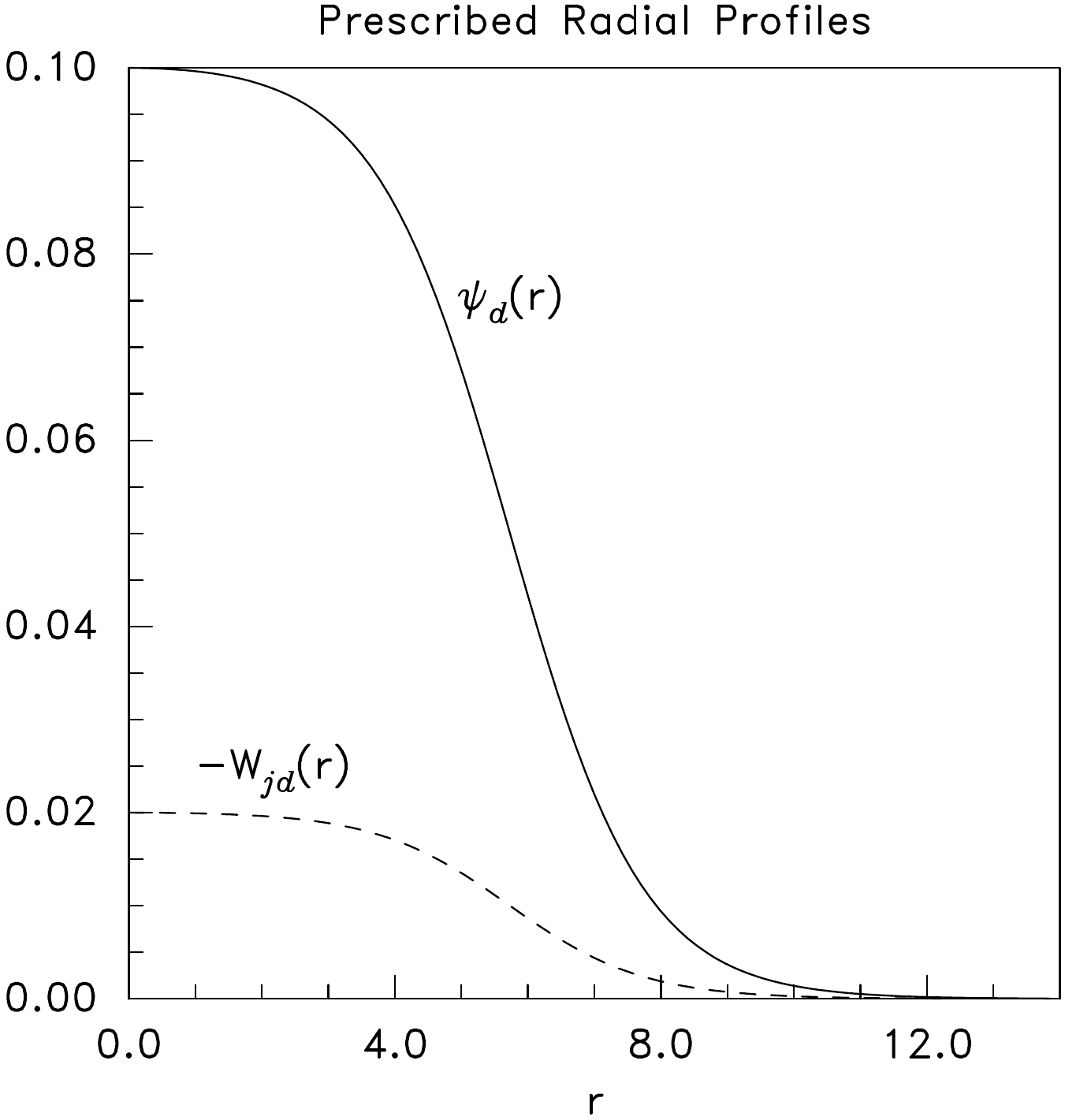}
  \caption{Example prescribed desired radial profiles of peak potential
    $\psi$ and negative join energy $-W_j$ ($\alpha=1/2$, $r_t=5$, $D=0$).}
  \label{fig:radialdesired1}
\end{figure}
Fig.\ \ref{fig:radialdesired1} shows an example of the prescribed
radial profiles of peak potential $\psi$ and negative join energy
$-W_j$ (for $\alpha=1/2$, $r_t=5$, $D=0$). 

\begin{figure}
  \centering
  \includegraphics[width=0.9\hsize]{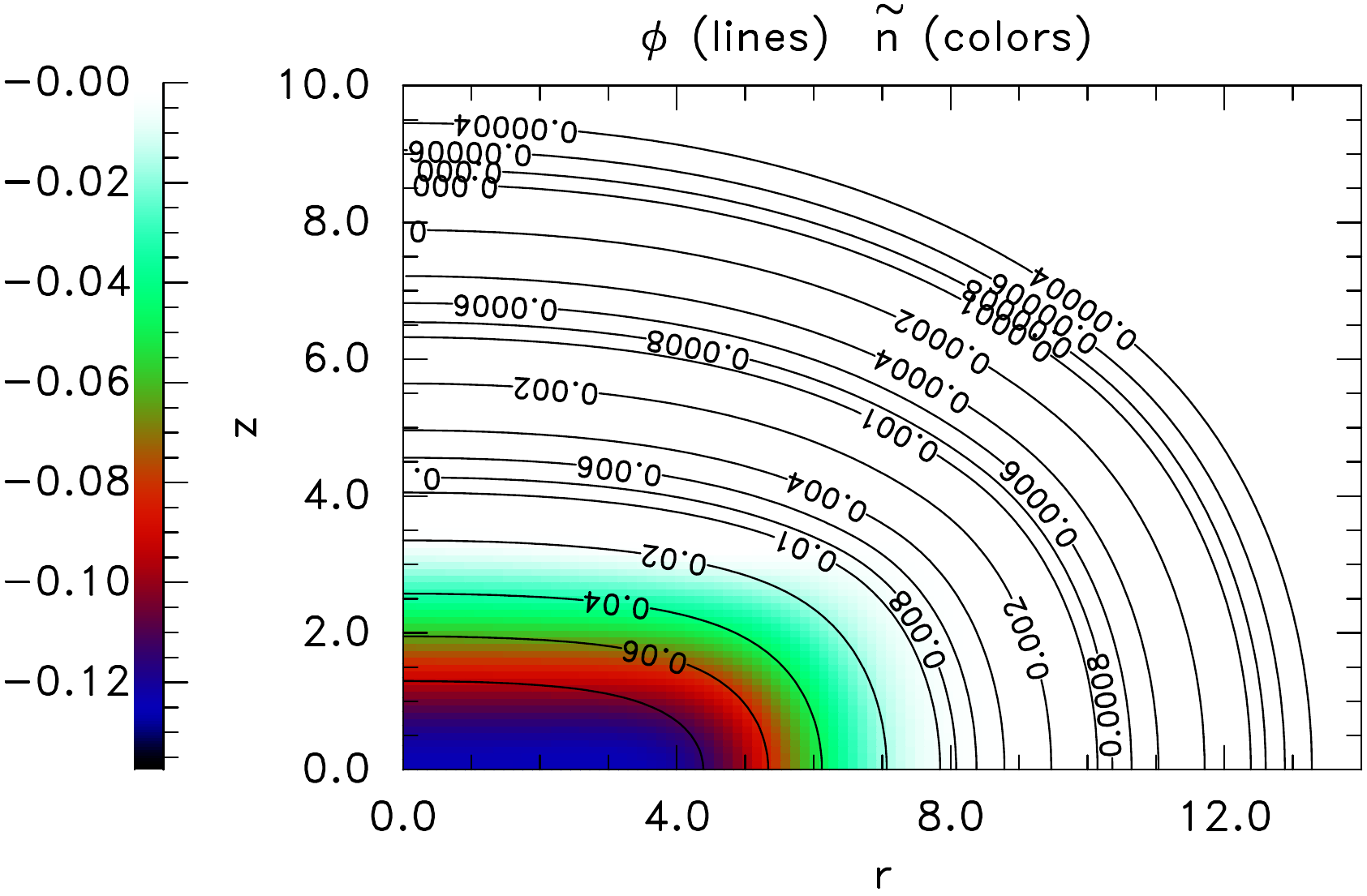}
  \caption{The final self-consistent hole solution based on the
    prescription of Fig.\ \ref{fig:radialdesired1}, showing
    logarithmically spaced contours of potential (lines) and the
    trapped-density deficit $\tilde n$ (colors) that sustains it.}
  \label{fig:phirho1}
\end{figure}

The final fully self-consistent equilibrium found for this example
(after 20 relaxation iterations) is shown in the contour plot of Fig.\
\ref{fig:phirho1}.
The chosen $r_t$ prescribes a hole considerably extended in the
transverse dimension. Beyond the region of non-zero electron density
deficit $\tilde n$, the potential contours gradually become less
oblate. The contours there show outward decay of potential that is
approximately exponential
${1\over\phi}{\partial \phi\over \partial R} \sim const.$; the spacing
of two adjacent contours is almost independent of angle in this
$rz$-plane. At radial positions $r\gtrsim r_t$ the contour lines are
nearly circular, concentric about $r\sim r_t$. Nearer the origin, the
potential contours are still oblate, but it is evident that they do
not follow lines of level density deficit $\tilde n$. For this
equilibrium, therefore, the deficit charge density
$\tilde\rho=-\tilde n$ is \emph{not} a function just of $\phi$. It is
a function also of $r$.  As has already been observed in the previous
section, such explicit $r$-dependence is essential to obtain
multidimensional (non-spherical) holes. That $r$-dependence is implicitly
prescribed by $\psi_d(r)$ and $W_{jd}(r)$. All these qualitative
comments are observed to apply to essentially the full range of
possible equilibria.

\begin{figure}
  \centering
  \includegraphics[width=0.8\hsize]{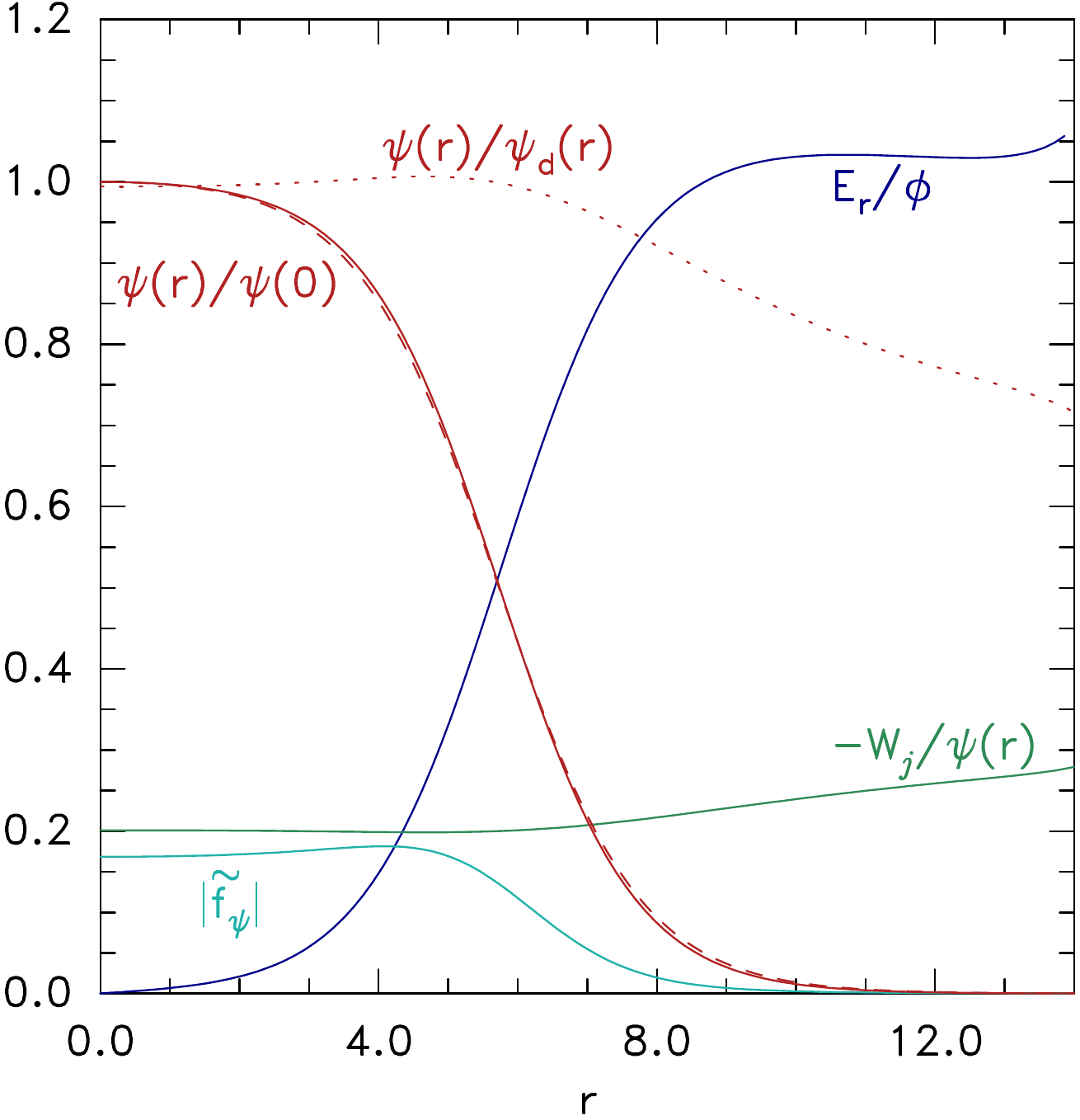}
  \caption{The radial shapes of the solved equilibrium parameters:
    potential $\psi(r)/\psi(0)$ with (dashed) its desired value
    $\psi_d(r)/\psi_d(0)$, and (dotted) their ratio, the peak distribution
    function deficit $|\tilde f_\psi|$, the ratio of join energy to potential
    peak $-W_j(r)/\psi(r)$, and the potential logarithmic gradient
    $E_r/\phi$ (at $z=0$).}
  \label{fig:solnrshapes1}
\end{figure}

In Fig. \ref{fig:solnrshapes1} are shown profiles of several
parameters along the radial axis ($z=0$). The solution potential
$\psi(r)$ normalized to its peak at the origin ($\psi(0)$) proves to
have a shape gratifyingly close to what was prescribed,
$\psi_d(r)/\psi_d(0)$ (the dashed line).
A way of showing the absolute agreement out to large radii is the
dotted line of $\psi(r)/\psi_d(r)$, which deviates significantly from
unity only at radii far from the hole, where the potential is already
small. The $z=0$ value of the distribution function deficit
$|\tilde f_\psi|$, which is fixed by its derivation prior to
relaxation from $\psi_d$ and $W_{jd}$, is slightly peaked off-axis (at
$r>0$). The reason is that it takes into account the transverse
electric field divergence $\nabla_\perp^2\phi$ of the prescribed
potential, which requires an enhanced deficit in regions of negative
$\psi$-curvature. The deficit $|\tilde f_\psi|$ falls quickly to zero
at large radius where the potential profile is governed essentially by
the screening effect of passing particles. If a potential that did not
have exponential radial variation in this distant region had been
prescribed, $|\tilde f_\psi|$ would not have become as quickly
negligible as it does here. This exponential potential variation at
distant $r$ is evident in the asymptotic approximately flat $E_r/\phi$
at a value of approximately 1. It is not exactly 1 because, although
$\lambda_s$ is effectively 1, the solution of the Helmholtz equation
is $\phi\propto {1\over r} \exp(-r/\lambda_s)$ with an extra $1/r$
factor contributing a correction of order $\lambda_s/r$ to the
slope.  The fractional join energy
$-W_j/\psi$ shows variation with radius that is the inverse of that of
$\psi(r)/\psi_d(r)$ because the iteration scheme holds the potential
difference $\psi-\phi_j=\psi_d-\phi_{jd}=\psi_d+W_{jd}$ fixed.

\begin{figure}
  \centering
  \includegraphics[width=0.9\hsize]{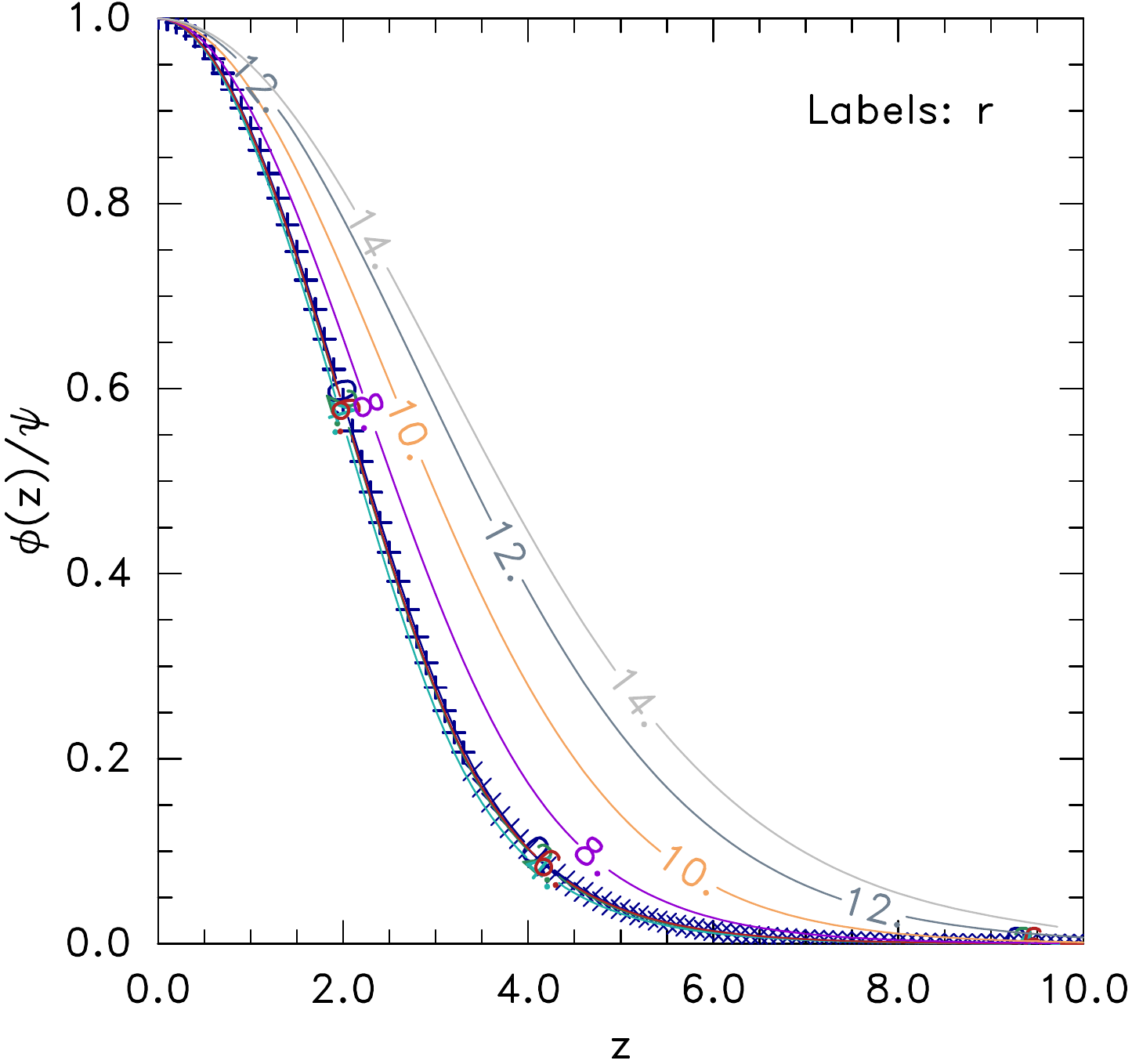}
  \caption{Parallel profiles of potential shape for a range of radial
    positions, showing the non-separable form of the solution. Marked
    points show the analytic power deficit model prescribed
    (`+' at $\phi_d>-W_j$ and `$\times$' at $\phi_d<-W_j$), in good
    agreement at small radius. }
  \label{fig:nonseparable1}
\end{figure}

Fig.\ \ref{fig:nonseparable1} shows at a range of $r$ positions (0, 2,
4, \dots) the relative \emph{parallel} shape $\phi(z)/\psi$. For
$r\lesssim r_t$ this shape is practically invariant. In this inner
radial range the potential is in fact separable to a good
approximation. And, as shown by the point markers plotted over the
lines, the agreement with the analytic solution for $\alpha=1/2$
(equations \ref{eq:cosform} and \ref{eq:expform}) in this region is
excellent.
At larger radii, however, the parallel extent of the potential profile
increases, showing that in the wing an assumption of separability is
inapplicable. 

\subsection{Consequences for satellite observations}
\begin{figure}
  (a)\hskip-1.5em\includegraphics[width=0.75\hsize]{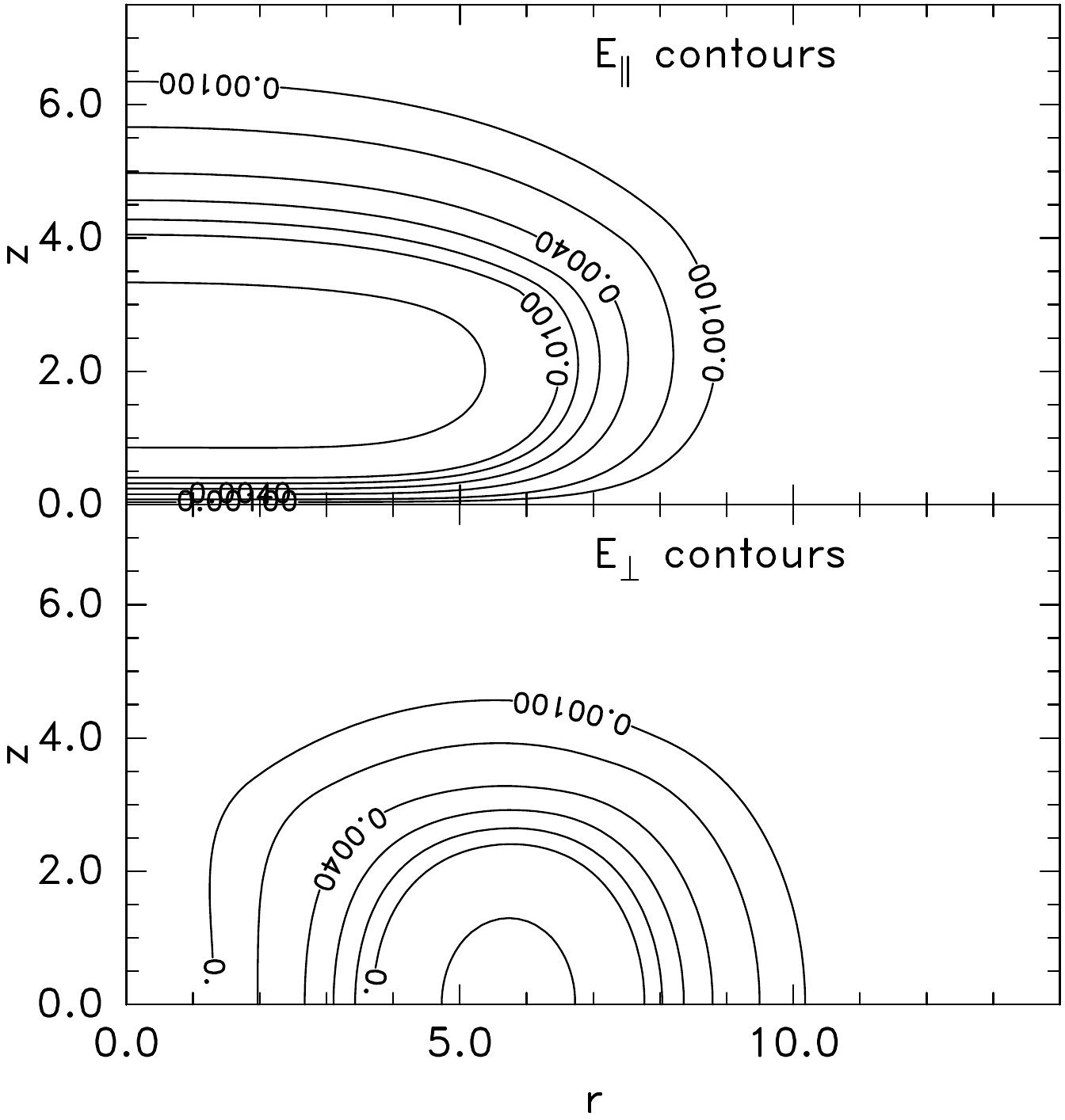}\phantom{MMM}

  \hskip1.em(b)\hskip-2.5em\includegraphics[width=0.9\hsize]{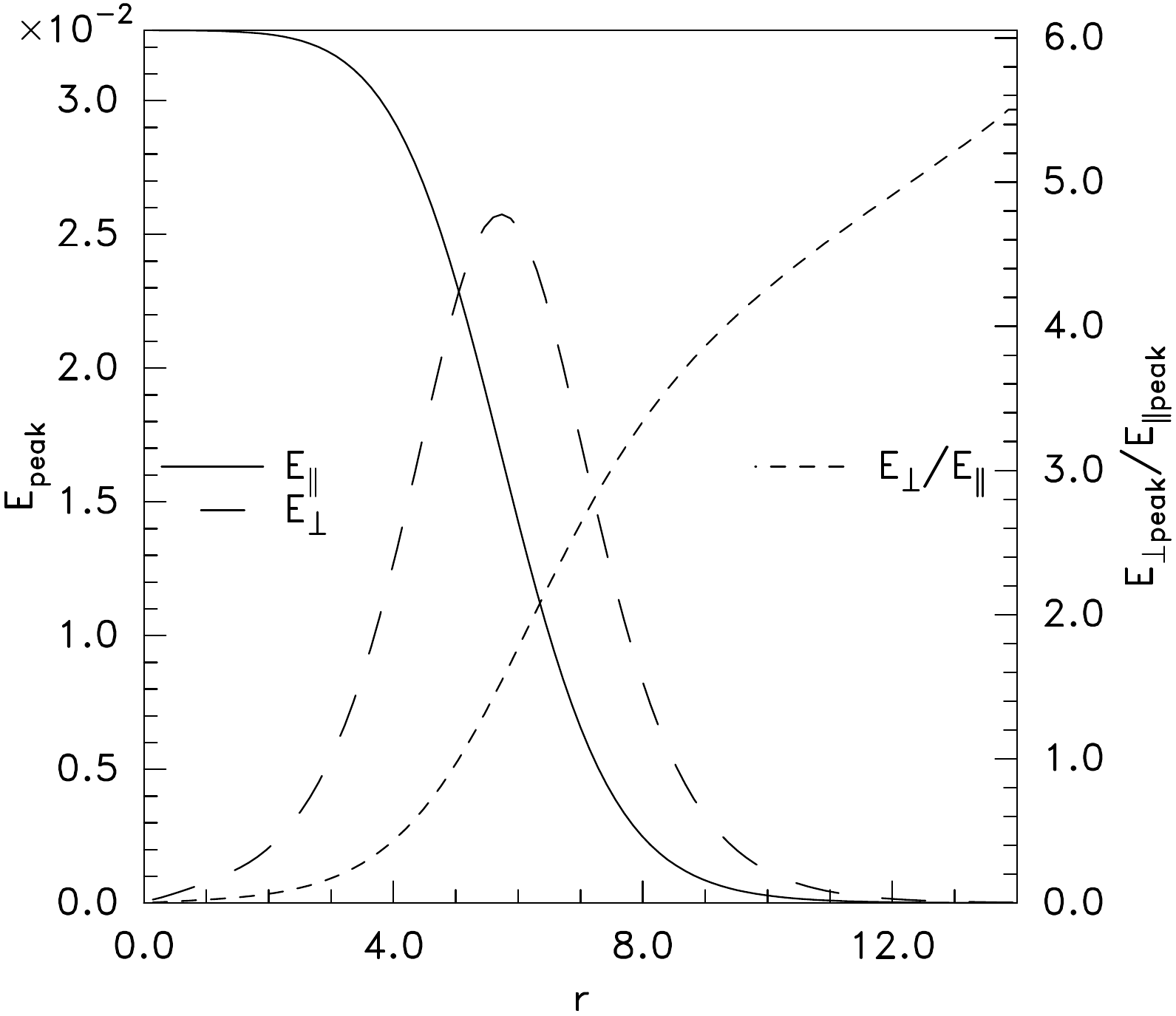}
  \caption{(a) Contours of parallel and perpendicular electric field
    corresponding to the equilibrium of Fig. \ref{fig:phirho1}. (b)
    Peak values of field along constant $r$ and their ratio versus
    radial position.}
  \label{fig:econtours1}
\end{figure}

Since satellites encountering electron holes measure the components of
the electric field, rather than the potential directly. It is helpful
to derive from an equilibrium like Fig. \ref{fig:phirho1} contours of
the perpendicular and parallel electric field. These are shown in
Fig.\ \ref{fig:econtours1}(a).
Generally holes move predominantly in the parallel direction at a
fraction of the electron thermal speed, which is usually much faster
than satellites move. Therefore to a first approximation,
the transit of an electron hole past a satelite corresponds to
relative motion of the satellite along a vertical (fixed-$r$) line
through the profile of the hole. The anticipated time
dependence of electric field is the corresponding spatial dependence
along such a line. As can easily be understood from the contours, such
motion gives rise to bipolar $E_\parallel$ and simultaneously unipolar
$E_\perp$ profiles. The occurrence of such features is often used to
identify electron hole events.  One simple and frequently used measure
of electron hole amplitude is the peak absolute value of the electric
field. For $E_\perp$ this usually occurs at $z=0$ the point of closest
approach; but for $E_\parallel$ it is always at
$|z|>0$. Fig. \ref{fig:econtours1}(b) therefore plots the peak values
of $|E_\perp|$ and $|E_\parallel|$ along each line $r=const$ as a function
of $r$.

The ratio of electric field peaks, $E_\perp/E_\parallel$, has often
been used in past satellite data analysis as a proxy for the aspect
ratio of an electron hole: that is, the ratio of the potential's scale
lengths in the parallel and perpendicular directions is supposed
$L_\perp/L_\parallel\sim
E_\parallel/E_\perp$. Fig. \ref{fig:econtours1} is an important
caution for this practice. The entire $r$-profile of this ratio comes
from a single electron hole model equilibrium, in which the actual
aspect ratio $L_\perp/L_\parallel$ of the density deficit is
approximately 3 and the aspect ratio of the potential contours
closer to 2. Nevertheless the model shows that the observed ratio
$E_\perp/E_\parallel$ can, all depending on $r$, be anywhere between 0
and $\sim5$. The \emph{most probable} or \emph{mean} value of the
ratio to be observed, supposing that the nearest distance of approach
is random, is determined by whatever selection criterion governs the
distant cut-off of selection of events as electron hole encounters. If
this cut-off is taken as sharp, corresponding to a specific radial
distance (e.g. $r=c$), or value of field component relative to its
peak (which would correspond to $r=c$), then the probability
distribution of detected radii is $p(r)=r/\int_0^cr dr=2r/c^2$ whose
maximum is at $r=c$ and whose mean is $r=2c/3$. The corresponding
probability distribution of $E_\perp(r)/E_\parallel(r)=A(r)$ is
\begin{equation}
  \label{eq:rprob}
  p(A)=p(r(A)){dr\over dA}={1\over c^2} {dr^2\over dA}. 
\end{equation}
The most probable value is $E_\perp/E_\parallel=A(c)$, and the mean
depends on the shape of the ratio profile but if $A(r)$ is
approximately parabolic it is $E_\perp/E_\parallel \sim A(c/2)$.  If
we suppose, for example, that the selection criterion gives $c=6$,
then for the profile shown we find the peak electric field ratio is
$\sim 1.8$ and the mean is $\sim 0.18$, while
if $c=8$ these values become $\sim3$ and $\sim0.25$. Thus, the most
probable field ratio gives naively an aspect ratio 4 to 6 times
smaller than the actual hole scale-length $L_\perp/L_\parallel$ and
the mean field ratio gives roughly twice the actual.  This is a
disappointingly large interpretational uncertainty. It reemphasizes
the statistical importance of the hole detection algorithm, and the
great value of obtaining multiple simultaneous satellite passages
through an electron hole (such as can be obtained using MMS\cite{Steinvall2019,Lotekar2020}) for
estimating the hole's transverse extent without relying on the
$E_\perp/E_\parallel$ ratio.

\subsection{Different radial shapes}
\begin{figure}
  (a)\hskip-1.5em
  \includegraphics[width=0.77\hsize]{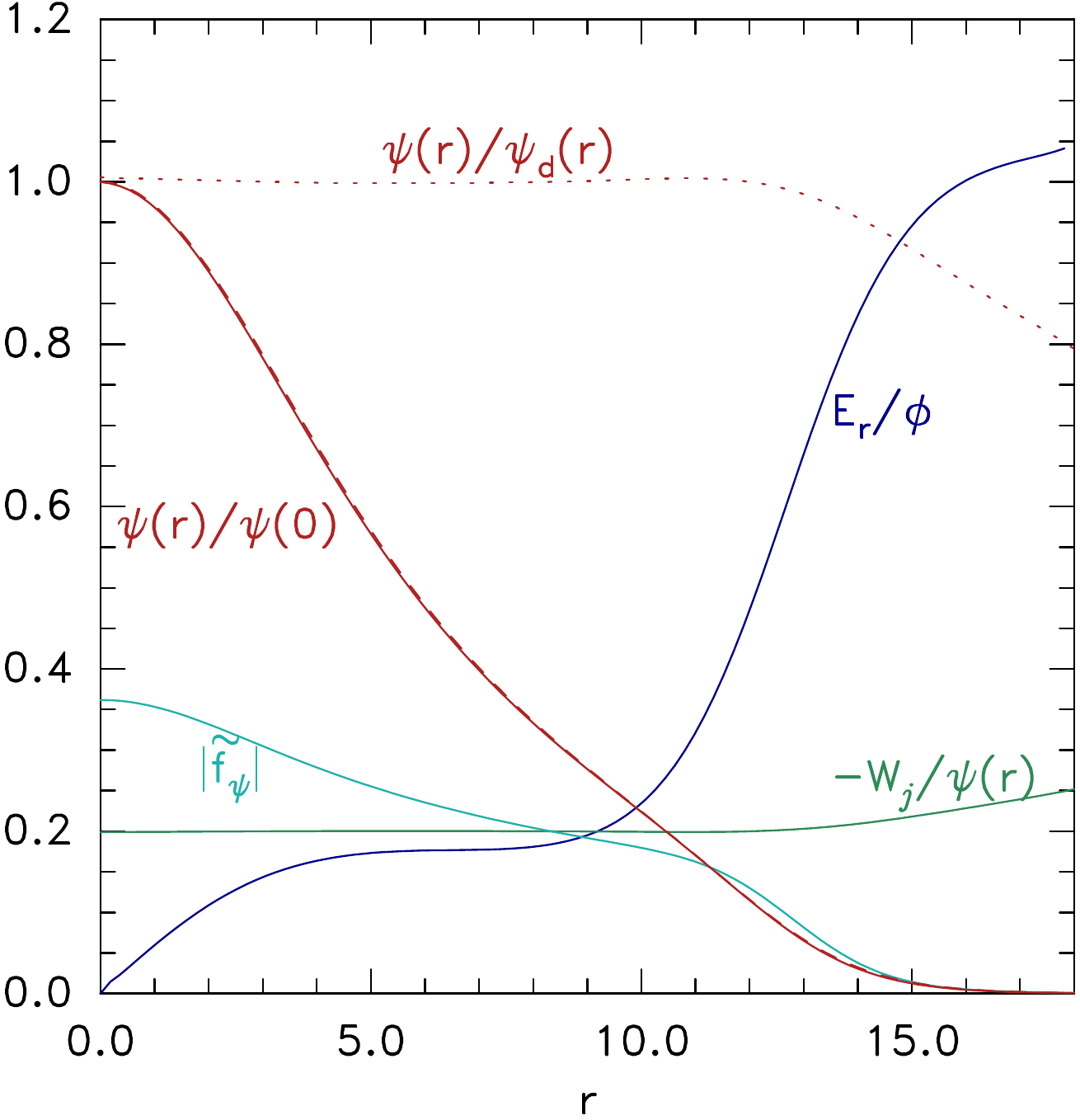}\phantom{MM}
  
  \hskip1.5em(b)\hskip-3em \includegraphics[width=0.95\hsize]{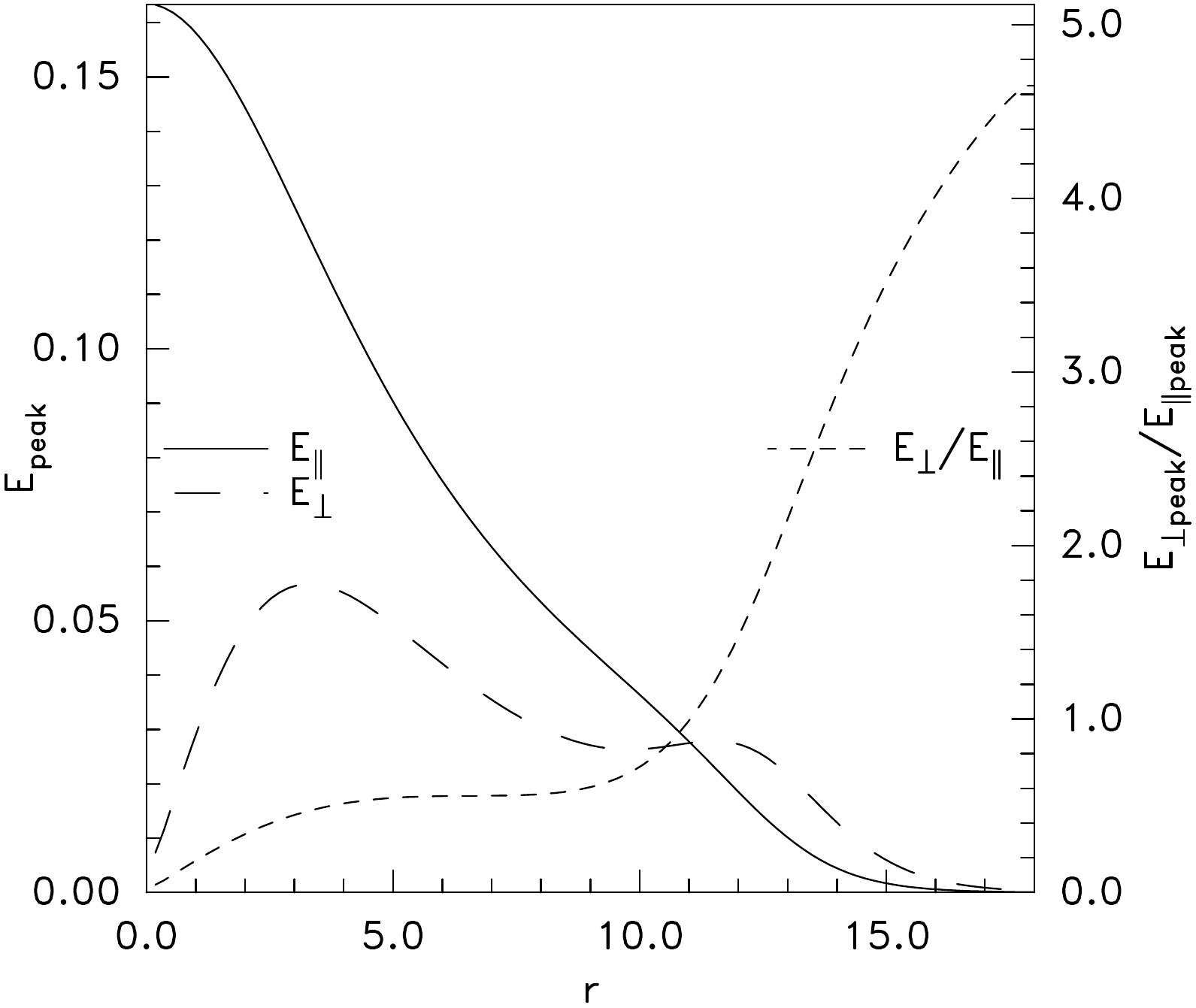}
  \caption{Equilibrium solution parameter radial profiles (a), and
    profiles of electric field peak values (b), for the hole
    illustrated in Fig.\ \ref{fig:onex}.\label{fig:onesum}}
\end{figure}

The earlier Fig.\ \ref{fig:onex} shown in the introduction is a
different example that illustrates the versatility of the equilibrium
prescription. The independent parameter adjustments make it more
oblate but with a more peaked radial profile ($r_t=10$, $D=0.03$,
$\alpha=1/2$), and greater amplitude ($\psi(0)=0.5$, $-W_j=0.1$).
Fig.\ \ref{fig:onesum} shows the summary plots, in which the shape changes
are evident.
The peak potential $\psi(r)$, in Fig.\ \ref{fig:onesum}(a), now has no
flat top, but falls steadily out to a radius of approximately $r=12$
under control of the more peaked $\tilde f_\psi$ profile, giving a
flatter $E_r$ profile. Then $\tilde f_\psi$ falls off rapidly into the
exponential region of $\psi(r)$ where $E_r\over \phi$ rises to unity.
Notice that the maximum value of $|\tilde f_\psi|\simeq 0.36$ is close
to the maximum permissible (0.399), showing that this equilibrium
requires the deeply trapped orbits to be almost completely depleted of
electrons near the origin. The peak fields along fixed $r$, Fig.\
\ref{fig:onesum}(b), show a more extended region of flat
$E_\perp/E_\parallel$ before it rises in the exponential
region. Still, even the flat region $E_\perp/E_\parallel\simeq 0.5$
does not agree very well with the typical inverse contour aspect ratio
$L_\parallel/L_\perp\sim 1/3$.  Despite the substantial radial
gradients, the parallel potential relative shapes (not shown) do not
significantly differ until $r\gtrsim 12$. They agree with the analytic
form in approximately the same way as Fig.\
\ref{fig:nonseparable1}. Thus, they are approximately separable until
a radius at which radial exponential decay sets in, like the previous
equilibrium example.

\section{Conclusions}

A technique for constructing fully self-consistent multidimensional
electron hole equilibria in the drift limit of negligible gyro-radius
has been described and illustrated. It starts from a versatile
phase-space distribution deficit form that satisfies the plausibility
constraint of being zero at and immediately below the parallel energy
trapping threshold, and from a desired radial transverse potential
variation. The analytic results of an approximation of separability of
the potential are derived. Dropping the separable approximation, which
cannot apply everywhere, requires an iterative relaxation to a fully
self-consistent (non-separable) equilibrium. Such relaxation cannot be
carried out under an assumption that the density deficit is a function
only of potential, because that leads to instability. An alternative
scheme, embodying numerically some aspects of the electron dynamic
response, finds multidimensional equilibria. But it also shows that no stable
non-spherical solitary equilibrium exists without there being explict
dependence of density deficit on transverse position. The deviation of
the full axisymmetric equilibria from the separable form proves to be
remarkably small except in the profile wings. The resulting electric
field component spatial dependencies, observable by satellites, are
illustrated for different oblate hole shapes. They show that the ratio
of peak perpendicular and parallel electric fields
$E_\perp/E_\parallel$ is unfortunately a very uncertain measure of the
aspect ratio $L_\parallel/L_\perp$ of the potential or charge
distributions of electron holes. This conclusion re-emphasizes the importance
of obtaining multiple simultaneous in-situ measurements such as can be
obtained from multiple-satellite missions, in order to establish the
true spatial structure of naturally occurring electron holes.

\subsection*{Supporting Data Statement}
The code used to calculate the equilibria and plot the figures of this
paper is open source and can be obtained from \href{https://github.com/ihutch/helmhole}{https://github.com/ihutch/helmhole}.

\bibliography{JabRef}

\end{document}